\documentclass[twocolumn,showpacs,prl,superscriptaddress]{revtex4}        
\def\pT{\mbox{$p_t $}} 
 
\def\sNN{\mbox{$\sqrt{s_{_{NN}}}$}} 
\newcommand{ \be }{\begin{equation}}     
\newcommand{ \ee }{\end{equation}}     
\newcommand{ \bea }{\begin{eqnarray}}     
\newcommand{ \eea }{\end{eqnarray}}     
\newcommand{ \la }{\langle}     
\newcommand{ \ra }{\rangle}

\usepackage{color}

\usepackage{graphicx} 
\usepackage{dcolumn} 
\usepackage{bm} 
\begin{document}        
\title{     
Azimuthal anisotropy and correlations 
at large transverse momenta in $p+p$  
and Au+Au collisions at \sNN = 200~GeV  
}
\affiliation{Argonne National Laboratory, Argonne, Illinois 60439}
\affiliation{University of Bern, 3012 Bern, Switzerland}
\affiliation{University of Birmingham, Birmingham, United Kingdom}
\affiliation{Brookhaven National Laboratory, Upton, New York 11973}
\affiliation{California Institute of Technology, Pasedena, California 91125}
\affiliation{University of California, Berkeley, California 94720}
\affiliation{University of California, Davis, California 95616}
\affiliation{University of California, Los Angeles, California 90095}
\affiliation{Carnegie Mellon University, Pittsburgh, Pennsylvania 15213}
\affiliation{Creighton University, Omaha, Nebraska 68178}
\affiliation{Nuclear Physics Institute AS CR, 250 68 \v{R}e\v{z}/Prague, Czech Republic}
\affiliation{Laboratory for High Energy (JINR), Dubna, Russia}
\affiliation{Particle Physics Laboratory (JINR), Dubna, Russia}
\affiliation{University of Frankfurt, Frankfurt, Germany}
\affiliation{Insitute  of Physics, Bhubaneswar 751005, India}
\affiliation{Indian Institute of Technology, Mumbai, India}
\affiliation{Indiana University, Bloomington, Indiana 47408}
\affiliation{Institut de Recherches Subatomiques, Strasbourg, France}
\affiliation{University of Jammu, Jammu 180001, India}
\affiliation{Kent State University, Kent, Ohio 44242}
\affiliation{Lawrence Berkeley National Laboratory, Berkeley, California 94720}
\affiliation{Massachusetts Institute of Technology, Cambridge, MA 02139-4307}
\affiliation{Max-Planck-Institut f\"ur Physik, Munich, Germany}
\affiliation{Michigan State University, East Lansing, Michigan 48824}
\affiliation{Moscow Engineering Physics Institute, Moscow Russia}
\affiliation{City College of New York, New York City, New York 10031}
\affiliation{NIKHEF, Amsterdam, The Netherlands}
\affiliation{Ohio State University, Columbus, Ohio 43210}
\affiliation{Panjab University, Chandigarh 160014, India}
\affiliation{Pennsylvania State University, University Park, Pennsylvania 16802}
\affiliation{Institute of High Energy Physics, Protvino, Russia}
\affiliation{Purdue University, West Lafayette, Indiana 47907}
\affiliation{University of Rajasthan, Jaipur 302004, India}
\affiliation{Rice University, Houston, Texas 77251}
\affiliation{Universidade de Sao Paulo, Sao Paulo, Brazil}
\affiliation{University of Science \& Technology of China, Anhui 230027, China}
\affiliation{Shanghai Institute of Applied Physics, Shanghai 201800, China}
\affiliation{SUBATECH, Nantes, France}
\affiliation{Texas A\&M University, College Station, Texas 77843}
\affiliation{University of Texas, Austin, Texas 78712}
\affiliation{Tsinghua University, Beijing 100084, China}
\affiliation{Valparaiso University, Valparaiso, Indiana 46383}
\affiliation{Variable Energy Cyclotron Centre, Kolkata 700064, India}
\affiliation{Warsaw University of Technology, Warsaw, Poland}
\affiliation{University of Washington, Seattle, Washington 98195}
\affiliation{Wayne State University, Detroit, Michigan 48201}
\affiliation{Institute of Particle Physics, CCNU (HZNU), Wuhan 430079, China}
\affiliation{Yale University, New Haven, Connecticut 06520}
\affiliation{University of Zagreb, Zagreb, HR-10002, Croatia}
\author{J.~Adams}\affiliation{University of Birmingham, Birmingham, United Kingdom}
\author{M.M.~Aggarwal}\affiliation{Panjab University, Chandigarh 160014, India}
\author{Z.~Ahammed}\affiliation{Variable Energy Cyclotron Centre, Kolkata 700064, India}
\author{J.~Amonett}\affiliation{Kent State University, Kent, Ohio 44242}
\author{B.D.~Anderson}\affiliation{Kent State University, Kent, Ohio 44242}
\author{D.~Arkhipkin}\affiliation{Particle Physics Laboratory (JINR), Dubna, Russia}
\author{G.S.~Averichev}\affiliation{Laboratory for High Energy (JINR), Dubna, Russia}
\author{S.K.~Badyal}\affiliation{University of Jammu, Jammu 180001, India}
\author{Y.~Bai}\affiliation{NIKHEF, Amsterdam, The Netherlands}
\author{J.~Balewski}\affiliation{Indiana University, Bloomington, Indiana 47408}
% ...
\collaboration{STAR Collaboration}\noaffiliation 
\begin{abstract}    
Results on high transverse momentum charged particle 
emission with respect to the reaction plane 
are presented for Au+Au collisions at \sNN= 200~GeV.
Two- and four-particle correlations results are presented as well as a
comparison of azimuthal correlations in Au+Au collisions to those in
$p+p$ at the same energy. Elliptic anisotropy, $v_2$, is found to reach its
maximum at $p_t \sim 3$~GeV/c, then decrease slowly and remain significant 
up to $p_t\approx 7$  -- 10 GeV/c.  Stronger suppression is found in the 
back-to-back high-$p_t$ particle correlations for particles emitted 
out-of-plane compared to those emitted in-plane.
The centrality dependence of $v_2$ at intermediate $p_t$
is compared to simple models based on jet quenching. 
\end{abstract}
\maketitle        

In high energy heavy-ion collisions, a high density system consisting of
deconfined quarks and gluons  is expected to be created~\cite{QGP}.
Energetic partons, resulting from initial hard
scatterings, are predicted to lose energy by induced gluon
radiation when propagating through the medium~\cite{quenching}. 
This energy loss is expected to depend
strongly on the color charge density of the created system and the
traversed path length of the propagating parton.
At Brookhaven's Relativistic Heavy Ion Collider (RHIC)
three different observations related to parton energy loss have
emerged: strong suppression of the inclusive hadron
production~\cite{Adcox:2002highpt130,Adler:2002xw,Adams:2003kv}, strong
suppression of the back-to-back high-$p_t$  jet-like
correlation~\cite{Adler:2002tq,Adams:2003im}, and large values
of the elliptic flow at high \pT~\cite{Adler:2002ct}. 
In non-central heavy ion collisions, the
geometrical overlap region has an almond shape in the transverse plane, 
with its short axis in the reaction plane. 
Depending on the emission azimuthal angle, 
partons traversing this system, on average, experience different path lengths 
and therefore different energy loss.
It leads to (a) azimuthal anisotropy in
high $p_t$ particle production 
with respect to the reaction plane~\cite{Snellings:1999gq,Wang:2000fq}
(the second harmonic in the particle azimuthal distribution, 
elliptic flow, is characterized~\cite{Posk98} by $v_2=\la
\cos2 ( \phi - \Psi_{R} ) \ra$.)
and (b) to the dependence of the high $p_t$ 2-particle 
back-to-back correlations on the orientation of the pair.

In this Letter, using
higher-order cumulant analysis~\cite{Olli00,Olli01} and 
comparing azimuthal correlations measured in $p+p$ collisions to
those in Au+Au, we confirm strong elliptic flow in mid-central Au+Au
collisions at least up to $p_t\approx7$~GeV/c
as qualitatively expected in the jet quenching scenario. 
We further investigate the influence of the jet quenching  
mechanism on high-$p_t$ particle production with respect to the 
reaction plane by studying $v_2$ 
centrality dependence in the intermediate $p_t$ region and 
two-particle azimuthal correlations 
at different angles with respect to the reaction plane.

The data set consists of about 2 million  
minimum bias and 1.2 million central trigger Au+Au events 
and 11 million $p+p$ events at \sNN= 200 GeV. 
The measurements were made using
the Time Projection Chamber~\cite{Anderson:2003ur} of the
STAR detector \cite{Ackermann:2002ad}, which covers 
pseudorapidity ($\eta$) from --1.3 to 1.3.
The event centrality in this paper is defined by the multiplicity 
measured at mid-rapidity by STAR~\cite{Adler:2002xw}.
Tracks used to reconstruct the flow vector, or generating 
function~\cite{Olli01} in the
case of the cumulant method, were subject to the same quality cuts as used 
in the \sNN = 130~GeV analysis~\cite{Ackermann:2000tr,Adler:2002pu}, 
except for the
low transverse momentum cutoff, which for this analysis is 0.15 GeV/c 
instead of 0.10 GeV/c. 

\begin{figure}[t]       
\vspace{-0.2cm}
  \includegraphics[width=0.50\textwidth]{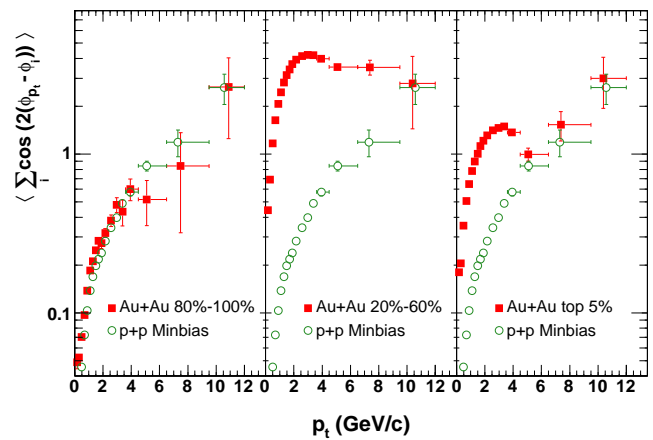}    
\vspace{-0.5cm}
  \caption{(color online)
    Azimuthal correlations in Au+Au collisions (squares) as a function
    of centrality (peripheral to central from left to right) compared to
    minimum bias azimuthal correlations in $p+p$ collisions (circles). Errors 
    are statistical only.}\vspace{-0.3cm}
  \label{fig:pp+AA-NewCent-NoWght}
\end{figure}

One of the largest uncertainties in elliptic flow measurements in nuclear
collisions is due to so-called non-flow 
effects -- the contribution to the
azimuthal correlations not related to the
reaction plane orientation, such as resonance decays 
and inter- and 
intra-jet correlations. 
The importance of these effects can be investigated by comparing
the azimuthal correlations measured in Au+Au to those
in $p+p$ collisions, where all
correlations are considered to be of non-flow origin.
For such a comparison we evaluate the accumulative 
correlation of a particle from a given $p_t$ bin with   
all other particles in the region $0.15<p_t<2.0$~GeV/c and
$|\eta|<1.0$ by calculating the event average sum: 
\be 
\la \sum_{i} \cos2(\phi_{p_t}-\phi_i) \ra = M \, v_{2}(p_t) \, 
\bar{v}_{2} 
+ \{ \text{non-flow}\} 
\label{eQAA}  
\ee 
where $\phi_{p_t}$ is the azimuthal angle of the particle 
from a given $p_t$ bin. 
The first term in the r.h.s. of Eq.~\ref{eQAA} 
represents the elliptic flow contribution,
where $v_2(p_t)$ is the elliptic flow of particles with a given $p_t$, and
$\bar{v}_2$ is the average flow of particles used in the sum;  
$M$ is the multiplicity of particles contributing
to the sum.
The multiplicity in the sum changes with the centrality of the
collision, but as long as the relative number of particles 
(per trigger particle) involved in non-flow effects does not change, the 
contribution due to these effects is a constant.
Comparing $p+p$ and Au+Au collisions one indeed might expect some
changes in particle correlations: there could be an increase in
correlations due to a possible increase 
of jet multiplicities in Au+Au collisions, or conversely,  
some decrease due to the
suppression of high $p_t$ back-to-back correlations~\cite{Adler:2002tq}.
It is difficult to make an accurate estimate of possible modifications
of non-flow effects. The fact that at very high $p_t$ the $p+p$ results are 
very close to central Au+Au (shown later by Figure~\ref{fig:pp+AA-NewCent-NoWght}), suggests that the modifications 
are relatively small.

Figure~\ref{fig:pp+AA-NewCent-NoWght} shows the azimuthal 
correlation, Eq.~\ref{eQAA}, as a
function of transverse momentum for 
three different centrality ranges in Au+Au collisions, as compared to
minimum bias $p+p$ collisions. 
We observe that for the most peripheral Au+Au collisions, 
the azimuthal correlations are very similar to minimum bias $p+p$. 
In mid-central Au+Au events, the azimuthal correlations 
are very different from those in $p+p$ collisions in both 
magnitude and $p_t$-dependence.
Note that at $p_t=7$~GeV/c, the azimuthal correlations 
in Au+Au collisions are
still many standard deviations away from those observed in $p+p$
collisions, indicating significant elliptic flow up to these momenta.
For the most central Au+Au collisions, at low-$p_t$ the magnitude of
the correlations is also different from $p+p$.  However, for particles
with $p_t \ge 5$~GeV/$c$, the correlation in Au+Au collisions starts 
to follow that in $p+p$ collisions,
suggesting that azimuthal correlations become dominated by
non-flow effects and that the latter are rather similar in $p+p$ 
and Au+Au collisions at those momenta.
The observed non-monotonic centrality
dependence of the azimuthal correlation at low and moderate $p_t$
is strong evidence of elliptic flow.
It is qualitatively different from 
that expected from intra-jet correlations among jet 
fragments~\cite{Yuri02}. 
\begin{figure}[t]   
  \includegraphics[width=0.47\textwidth]{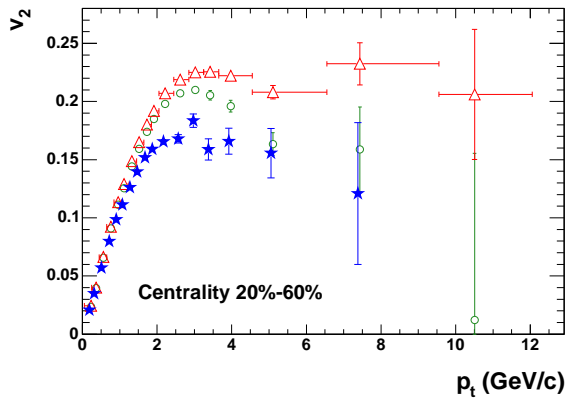}    
\vspace{-0.2cm}
  \caption{(color online) 
    $v_2$ of charged particles 
	as a function of transverse momentum from 
    the two-particle cumulant method (triangles) and four-particle
    cumulant method (stars). 
Open circles show the 2-particle
correlation results 
after subtracting the correlations 
measured in $p+p$ collisions. Only statistical errors are shown.  
  }\vspace{-0.3cm}
  \label{fig:v2Pt}
\end{figure}
        
We also perform a
multi-particle cumulant analysis, which is much less sensitive to 
non-flow effects than the traditional approach based on two-particle
correlations.
Figure~\ref{fig:v2Pt} shows $v_2$ as a function of
transverse momentum for 20\%-60\% of the total cross-section. 
The $v_2$ obtained using the four-particle cumulant method, $v_2\{4\}$, 
is up to about 20\% lower than the value of $v_2$ obtained from the two-particle
cumulant method.
This difference could be partially explained by non-flow
effects, which are greatly suppressed in $v_2\{4\}$, and by the fluctuation of 
$v_2$ itself~\cite{Adler:2002pu,Miller:2003kd}. Flow fluctuations contribute
to $v_2\{2\}$ and $v_2\{4\}$ with different signs. The true $v_2$ lies  
between $v_2\{4\}$ and approximately the average of $v_2\{2\}$ and $v_2\{4\}$.
The systematic uncertainty is given by these two bounds. 
For the centrality range plotted in Fig.~\ref{fig:v2Pt}, we find
significant $v_2$ at least up to $p_t\approx 7$~GeV/c,
well within the region where particle production is expected 
to be dominated by parton fragmentation. 
Two-particle cumulant results extend to 12~GeV/c, 
although at high $p_t$ these might be dominated by non-flow contributions. 
Also shown in Fig.~\ref{fig:v2Pt} by open circles are the 2-particle
correlation results 
after subtracting the correlations 
measured in $p+p$ collisions.
The comparison of these results to $v_2\{4\}$ in the region  $p_t<4$~GeV/c
indicates that either the relative
contribution of non-flow effects is larger in Au+Au collisions
compared to $p+p$, or there is a significant flow fluctuation
contribution that would increase the apparent $v_2\{2\}$ values 
and decrease $v_2\{4\}$.
In general we observe that $v_2(p_t)$ reaches a maximum at about
3~GeV/c,
confirming results obtained by PHENIX~\cite{v2phenix2003} 
and then slowly decreases.

The energy loss mechanism that leads to azimuthal anisotropy at 
high $p_t$ also leads to a distinct feature in two particle azimuthal
correlations. At high transverse momenta,
two-particle distributions in the relative azimuthal angle 
measured in $p+p$, $d$+Au, and Au+Au collisions 
at RHIC \cite{Adler:2002ct,Adler:2002tq,Adams:2003im} exhibit a jet-like 
correlation characterized by the 
peaks at $\Delta\phi=0$ (near-side correlations) 
and at $\Delta\phi=\pi$ (back-to-back).  
The back-to-back peak is
found to be strongly suppressed in central Au+Au collisions
\cite{Adler:2002tq}.
In non-central collisions, the suppression should depend on the relative
orientation of the back-to-back pair with respect to the reaction plane.
In the analysis of the two-particle azimuthal correlations,
we select {\it trigger} particles with 
$4<p_t^{\rm trig}<6$~GeV/c emitted in the direction of the event 
 plane angle $\Psi_2$ 
(in-plane, $|\phi^{\rm trig}-\Psi_2|<\pi/4$  
and $|\phi^{\rm trig}-\Psi_2|>3\pi/4$) and perpendicular to it 
(out-of-plane, $\pi/4<|\phi^{\rm trig}-\Psi_2|<3\pi/4$). 
The trigger particles are paired with {\it associated} 
particles satisfying 2~GeV/c$ <p_t<p_t^{\rm trig}$. 
The tracks are restricted to $|\eta|<$ 1. 
To reduce the effect of particles produced within a jet on 
the reaction plane reconstruction,
all particles in a pseudorapidity 
region $|\Delta\eta|<0.5$ around the 
highest $p_t$ particle
in the event are excluded from the event plane determination.
\begin{figure}[t]       
\vspace{-0.5cm}
\resizebox{20pc}{!}{\includegraphics{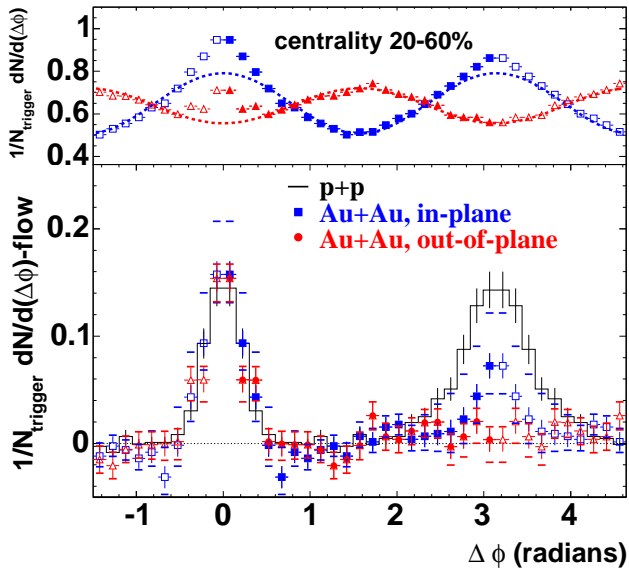}}       
\vspace{-2cm} 
\caption{\label{fig:42bins} (color online) 
Upper panel: Azimuthal distributions of 
associated particles
for trigger particles in-plane (squares) and out-of-plane (triangles)  
for Au+Au collisions at centrality 20-60\%.  
Open symbols are reflections of solid symbols  
around $\Delta\phi=0$ and $\Delta\phi=\pi$.  
Elliptic flow contribution is shown by dashed lines. 
Lower panel: Distributions after substracting elliptic flow, and the 
corresponding measurement in $p+p$ collisions (histogram).
} \vspace{-0.3cm}
\end{figure}        
In the upper panel of Fig.~\ref{fig:42bins} we show the azimuthal 
distributions of associated particles for trigger particles that 
are in-plane (squares) 
and out-of-plane (triangles) in midcentral
Au+Au collisions. 
The distributions are corrected for the reconstruction efficiency.  
The measured distributions exhibit a strong elliptic flow pattern 
similar to that found in the recent analysis at the SPS \cite{CERES}.

In the presence of elliptic flow the  
in-plane and out-of-plane two-particle azimuthal 
distributions are given by \cite{2partRP}: 
\begin{eqnarray} 
 \frac{d{\rm n_{out}^{in}
}}{d\Delta\phi}& = 
B\left[1+2v_2^{\rm assoc}
\left(
\frac{\pi v_2^{\rm trig} \pm 2\langle\cos{(2\Delta\Psi)}\rangle}
{\pi \pm 4v_2^{\rm
trig}\langle\cos{(2\Delta\Psi)}\rangle}
\right)
\cos{(2\Delta\phi)}
\right], 
\label{eq:inout} 
\end{eqnarray} 
where $v_2^{\rm assoc}$ and $v_2^{\rm trig}$ are the elliptic flow 
of the associated and trigger particles, respectively, and
$\langle\cos{2\Delta\Psi}\rangle$ is the reaction plane  
resolution~\cite{Posk98}. 
For the given centrality $\langle\cos{(2\Delta\Psi)}\rangle=0.70$;
 $v_2^{\rm assoc}=0.20$, and $v_2^{\rm trig}=0.18$ 
measured  via the reaction plane method. 
For the estimate of the systematic uncertainty in the 
determination of the flow contribution, we have varied $v_2^{\rm assoc}$ and 
$v_2^{\rm trig}$ between 0.167 and 0.213 ($v_2\{4\}$ and $v_2\{2\}$ measured in the range
2$<p_t<$6 GeV/c). To reduce the systematics, a $z$ vertex cut of $\pm 25$ cm 
is applied to $p+p$ events to match that in $Au+Au$ events.

The distributions were fit to Eq.~\ref{eq:inout}
in the region $0.75<|\Delta\phi|<2.24$ rad, with $B$ as the only  
free parameter, to determine the amount of background. 
For the in-plane distribution, $B=0.649 \pm 0.004(stat.) \pm 0.005(sys.)$, 
and for the out-of-plane, $B=0.638 \pm 0.004(stat.) \pm 0.002(sys.)$.
The systematic errors were estimated from using different ranges of $\Delta\phi$ in the fit.
We observe a strong excess of two-particle correlations over the correlation
pattern generated by elliptic flow  
in the region $|\Delta\phi|<0.75$ for 
both in-plane and out-of-plane distributions, characteristic of 
near-side intra-jet correlations.  
In the region around $\Delta\phi=\pi$, we observe an excess 
for the in-plane distribution, 
but no excess is found for the out-of-plane distribution.  
This is better illustrated in the lower 
panel of Fig.~\ref{fig:42bins}, where we show the flow-subtracted 
in-plane and out-of-plane distributions compared to that measured 
in $p+p$ collisions. The level of combinatorial 
background measured in $p+p$ collisions, $0.014\pm 0.001$, 
has been subtracted.
The near-side jet-like correlations measured in Au+Au 
are similar to those measured  in $p+p$ collisions. 
The back-to-back (around $\Delta\phi=\pi$)  
correlations measured in Au+Au collisions for in-plane trigger 
particles are suppressed compared to $p+p$, and even more suppressed 
for the out-of-plane trigger particles. 
For the near angle correlations in the relative azimuthal region
$|\Delta \phi |<0.75$ rad, the integrals of the azimuthal
distributions are $0.078 \pm 0.014(stat.)^{+0.059}_{-0.031}(sys.)$ in-plane and 
$0.081 \pm 0.014(stat.)^{+0.004}_{-0.021}(sys.)$ out-of-plane.  
 For the back-to-back correlations in the relative azimuthal region 
$|\Delta \phi - \pi|<0.75$ rad, the integrals are 
$0.048 \pm 0.014(stat.)^{+0.059}_{-0.031}(sys.)$ in-plane and 
$0.014 \pm 0.014(stat.)^{+0.004}_{-0.021}(sys.)$ 
out-of-plane. 
Note that the large systematic errors in Fig.~\ref{fig:42bins} (lower panel), 
resulting from the uncertainty 
in the subtraction of elliptic flow contribution, are  
highly anti-correlated: assuming weaker (stronger) elliptic flow results in 
the upper (lower) systematic error bar for 
$d{\rm n^{in}}/d\Delta\phi$ 
and lower (upper) systematic error bar for 
$d{\rm n^{out}}/d\Delta\phi$ distributions.

A different approach to remove the elliptic flow 
contribution to the two-particle distributions is 
to subtract the raw away-side
correlations from the near-side correlations measured 
in the same $|\Delta \phi|$ range (in this case, the elliptic flow
contribution cancels out).
The difference in the correlation strength, 
an integral over $\Delta \phi$ region, 
on the near-side ($|\Delta \phi|<0.75$ rad) and the away-side
($|\Delta \phi-\pi|<0.75$ rad) is measured to be 
$0.030 \pm 0.011(stat.)$
for the in-plane triggers and $0.067 \pm 0.011(stat.)$ 
for the out-of-plane triggers
where the systematic uncertainty due to elliptic flow 
is canceled out, and the remaining 
systematic uncertainties are smaller than the statistical errors.
Assuming similar strength of the near-side correlations in-plane and 
out-of-plane, the observed difference can be attributed to the suppression
of away-side correlations which depends on the reaction plane orientation. 

Although results presented above strongly support the
jet-quenching scenario qualitatively,
 the amount of elliptic flow observed at high $p_t$ for collisions at
\sNN = 130 GeV seems to exceed the values expected in the case of
complete quenching~\cite{Shuryak02}. 
Extreme quenching leads to emission of high-$p_t$ particles
predominantly from the surface, and in this 
case $v_2$ would be fully determined
by the geometry of the collision. 
This hypothesis can be tested by studying 
the centrality dependence of $v_2$ for high-$p_t$ particles.
\begin{figure}[t]       
\vspace{-0.5cm}
  \includegraphics[width=0.5\textwidth]{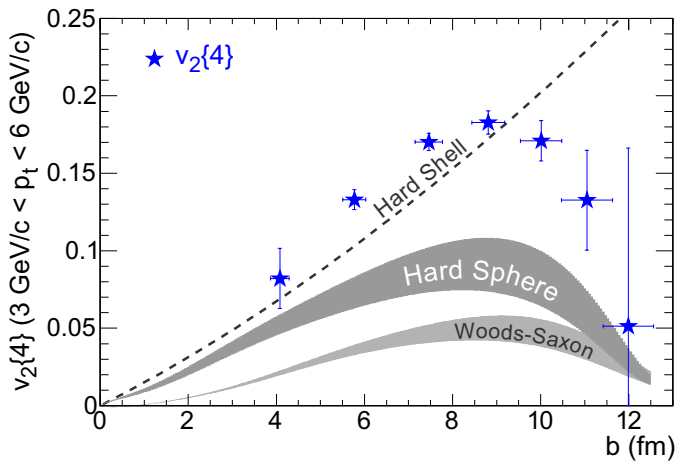}    
\vspace{-0.5cm}
  \caption{(color online)  
    $v_2$ at $3 \le p_t \le 6$ GeV/$c$ versus impact parameter, 
      $b$, compared to models of particle emission by a static source
    (see text).
  }   
  \label{fig:v2SurfaceEmission_b}\vspace{-0.3cm}
\end{figure}

Figure~\ref{fig:v2SurfaceEmission_b} shows 
$v_2$ in the $p_t$-range of 3--6 GeV/$c$ (where $v_2$ is
approximately maximal and constant) versus impact parameter. 
The values of the impact parameters were obtained using 
a Monte Carlo Glauber calculation~\cite{Glauber}.  
The measured values of $v_2\{4\}$ are compared to various
simple models of jet quenching. 
The upper curve corresponds to a complete  
 quenching, in which particles are emitted 
from a hard shell~\cite{SergeiQM02,Shuryak02};
this gives the maximum values of $v_2$ that are
possible in a surface emission scenario.
A more realistic calculation corresponds to a parameterization of jet
energy loss in a static medium where the absorption coefficient is set to
match the suppression of the inclusive hadron yields~\cite{Adams:2003kv}. 
The density distributions of the static medium are modeled using a
step function (following~\cite{XinNianJetQuenching03}) and a more 
realistic Woods-Saxon
distribution (following~\cite{Axel03}). The corresponding $v_2$ values are
shown as the upper and lower band, respectively. 
The lower and 
upper boundaries of bands correspond to an absorption 
that gives a suppression factor of 3 and 5~\cite{Adams:2003kv}, respectively, in central collisions.
Over the whole centrality range,  the measured $v_2$ values 
are much larger compared to calculations.  
Taking into account that this measurement is dominated by the
lower $p_t$ side ($3$ GeV/c),
the quark coalescence mechanism~\cite{q-coalescence} might 
be responsible for the difference, but no quantitative
explanation for the observed large elliptic flow exists at the moment. 

In summary, we have shown that the charged particle elliptic anisotropy 
in midcentral Au+Au collisions at $\sNN$ =200~GeV extends 
to large transverse momenta, at least up to $p_t\sim 7$~GeV/c,
as expected in a jet quenching scenario.  
By performing multi-particle correlation analysis and
comparing the azimuthal correlations in Au+Au collisions  
to those in $p+p$, we find the contribution of the effects
not associated with the reaction plane orientation  
is relatively small in midcentral events  
but  could be significant in peripheral and central collisions.  
We report stronger  suppression  
of the back-to-back high $p_t$ correlations  
for out-of-plane triggers 
compared to in-plane triggers, again  
consistent with a jet quenching picture. 
$v_2$ integrated from moderate to high $p_t$,
approximately in the region where it reaches a maximum, 
clearly exceeds the limits set for elliptic flow due to a simple jet
quenching mechanism, and still waits for quantitative theoretical
explanation.

\begin{acknowledgments}        
We thank the RHIC Operations Group and RCF at BNL, and the
NERSC Center at LBNL for their support. This work was supported
in part by the HENP Divisions of the Office of Science of the U.S.
DOE; the U.S. NSF; the BMBF of Germany; IN2P3, RA, RPL, and
EMN of France; EPSRC of the United Kingdom; FAPESP of Brazil;
the Russian Ministry of Science and Technology; the Ministry of
Education and the NNSFC of China; SFOM of the Czech Republic,
FOM and UU of the Netherlands,
DAE, DST, and CSIR of the Government of India; the Swiss NSF.
\end{acknowledgments}

\end{document}